\documentclass{ws-mpla}
\newcommand{\be}{\begin{equation}}
\newcommand{\ee}{\end{equation}}
\newcommand{\bea}{\begin{eqnarray}}
\newcommand{\eea}{\end{eqnarray}}
\newcommand{\ba}{\begin{array}}
\newcommand{\ea}{\end{array}}

\begin{document}

\markboth{Harleen Dahiya and Neetika Sharma} {Strangeness and
chiral symmetry breaking}

\catchline{}{}{}{}{}

\title{STRANGENESS AND CHIRAL SYMMETRY BREAKING }

\author{\footnotesize HARLEEN DAHIYA\footnote{
Permanent address.}}
\address{Department of Physics, Dr. B.R. Ambedkar National
Institute of Technology,\\ Jalandhar, 144011, India\\
dahiyah@nitj.ac.in}

\author{NEETIKA SHARMA}
\address{Department of Physics, Dr. B.R. Ambedkar National
Institute of Technology,\\ Jalandhar, 144011, India}

\maketitle


\begin{abstract}

The implications of chiral symmetry breaking and SU(3) symmetry
breaking have been studied in the chiral constituent quark model
($\chi$CQM). The role of hidden strangeness component has been
investigated for the scalar matrix elements of the nucleon with an
emphasis on the meson-nucleon sigma terms. The $\chi$CQM is able
to give a qualitative and quantitative description of the ``quark
sea'' generation through chiral symmetry breaking. The significant
contribution of the strangeness is consistent with the recent
available experimental observations.
\end{abstract}
\ccode{PACS Nos.: 12.39.Fe, 14.20.-c, 13.75.-n}

The internal structure of the nucleon has been extensively studied
over the past 40 or 50 years and it is still a big challenge to
perform the calculations from the first principles of Quantum
Chromodynamics (QCD).  The measurements of polarized structure
functions of proton in the deep inelastic scattering (DIS)
experiments\cite{emc1,emc2,smc,adams1,adams2,adams3,adams4,hermes}
provided the first evidence that the valence quarks of proton
carry only a small fraction of its spin suggesting that they
should be surrounded by an indistinct sea of quark-antiquark
pairs. These observations were in contradiction with the
predictions of Naive Quark Model
(NQM)\cite{nqm1,nqm2,nqm3,nqm4,nqm5} which is able to provide a
intuitive picture of the nucleon and successfully accounts for
many of the low-energy properties of the hadrons in terms of the
valence quarks.

Several interesting facts have also been revealed regarding the
flavor distribution functions in the famous New Muon
Collaboration\cite{nmc1,nmc2} and E866
experiments\cite{e8661,e8662,e8663} indicating that the flavor
structure of the nucleon is not limited to $u$ and $d$ quarks
only. The measured quark sea asymmetry of the unpolarized quarks
in the nucleon established that the study of the structure of the
nucleon is intrinsically a nonperturbative phenomena and is
considered as one of the most
active areas in the present day. 

Recently, there have been indications of strangeness contribution
in the experiments measuring electromagnetic form factors, for
example, SAMPLE at MIT-Bates\cite{sample}, G0 at JLab\cite{g0}, A4
at MAMI\cite{a4} and HAPPEX  at JLab\cite{happex1,happex2}. These
experiments have provided considerable insight on the role played
by strange quarks when the nucleon interacts at high energies. On
the other hand, a non-zero strangeness content in the nucleon
$y_N$ has been indicated in the context of low-energy
experiments\cite{nutev,pion nucleon1,pion nucleon2,pion
nucleon3,pion nucleon4}. Even though there has been considerable
progress in the past few years to estimate the strangeness matrix
elements from the neutral current observables, there is no
consensus regarding the various mechanisms which can contribute to
$y_N$\cite{gasser1,gasser2,riaz}. Since the strange quarks
constitute purely sea degrees of freedom, the low-energy
determination of the strangeness contribution to the nucleon would
undoubtedly provide vital clues to the nonperturbative aspects of
QCD.

Currently, there is enormous interest in determining the
meson-nucleon sigma terms\cite{pion nucleon1,pion nucleon2,pion
nucleon3,pion nucleon4}. These are the fundamental parameters to
test the chiral symmetry breaking ($\chi$SB) effects and thereby
determine the scalar quark content of the baryons.  The
meson-nucleon sigma terms cannot be measured directly from
experiments and are known to have intimate connection with the
dynamics of the non-valence quarks. They are theoretically
interesting because there is a discrepancy in the value derived
from the meson-nucleon scattering
experiments\cite{koch1,koch2,koch3,Gasser:1990ce1,Gasser:1990ce2,Pavan,hite}
and from the hadron spectroscopy\cite{riaz,cheng rev}. The
meson-nucleon sigma terms also provide restriction on the
contribution of strangeness to the parameters measured in
low-energy\cite{bass1,bass2,bass3,bass4,bass5,bass6}.

One of the most successful model which can yield an adequate
description in this energy regime is the chiral constituent quark
model ($\chi$CQM)\cite{manohar1,manohar2,manohar3}. The $\chi$CQM
is not only successful in giving a satisfactory explanation of
``proton spin crisis''\cite{hd1,hd2,hd3,hd4,hd5}, baryon magnetic
moments\cite{hdmagnetic1,hdmagnetic2} and hyperon $\beta-$decay
parameters\cite{nsweak1,nsweak2} but is also able to account for
the violation of Gottfried Sum
Rule\cite{gsr,hdasymmetry1,hdasymmetry2} and Coleman-Glashow sum
rule\cite{hdmagnetic1,hdmagnetic2,cgsr}. Recently, the
comparatively large masses of the strange quarks has been
reiterated in detail through SU(3) symmetry
breaking\cite{hd1,hd2,hd3,hd4,hd5,nsweak1,nsweak2} and the
predictions are found to improve in the case of spin polarization
functions and related parameters. In this context, it therefore
becomes desirable to carry out a detailed analysis of the role
played by chiral symmetry breaking and SU(3) symmetry breaking in
understanding the dynamics of quark sea in the nonperturbative
regime of QCD with an emphasis on the strangeness flavor
distribution functions.

The purpose of the present communication is to understand the
implications of chiral symmetry breaking ($\chi$SB) for the scalar
matrix elements of the nucleon within the $\chi$CQM. In
particular, we would like to phenomenologically estimate the
quantities affected by the hidden strangeness component in the
nucleon, for example, strangeness content in the nucleon $y_N$ and
strangeness fraction $f_s$. Further, it would be significant to
study the meson-nucleon sigma terms ($\sigma_{KN}$, $\sigma_{\eta
N}$) which have not been observed experimentally and are expected
to have large contributions from the quark sea. Furthermore, it
would be interesting to understand the extent to which the strange
quark mass contribute through the SU(3) symmetry breaking effects
in understanding the nucleon properties.

For ready reference as well as to make the mss. more readable,  we
present the essentials of $\chi$CQM. The key to understand the
structure of the nucleon, in the $\chi$CQM
formalism\cite{cheng1,cheng2,cheng3,cheng4}, is the fluctuation
process \be q^{\pm} \rightarrow {\rm GB} + q^{' \mp} \rightarrow
(q \bar q^{'}) +q^{'\mp}\,, \label{basic} \ee where GB represents
the Goldstone boson and $q \bar q^{'} +q^{'}$ constitute the
``quark
sea''\cite{hd1,hd2,hd3,hd4,hd5,cheng1,cheng2,cheng3,cheng4,johan1,johan2}.
The effective Lagrangian describing the interaction between quarks
and a nonet of GBs,  can be expressed as \be {\cal L}= g_8 {\bf
\bar q}\left(\Phi+\zeta\frac{\eta'}{\sqrt 3}I \right) {\bf q}=g_8
{\bf \bar q}\left(\Phi' \right) {\bf q}\,,\label{lag} \ee where
$\zeta=g_1/g_8$, $g_1$ and $g_8$ are the coupling constants for
the singlet and octet GBs, respectively, $I$ is the $3\times 3$
identity matrix. The parameter $a(=|g_8|^2$) denotes the
probability of chiral fluctuation  $u(d) \rightarrow d(u) +
\pi^{+(-)}$. The SU(3) symmetry breaking parameters $\alpha$,
$\beta$ and $\zeta$ are introduced by considering $M_s > M_{u,d}$,
$M_{K,\eta}> M_{\pi}$ and $M_{\eta^{'}} >
M_{K,\eta}$\cite{cheng1,cheng2,cheng3,cheng4,johan1,johan2} and
they respectively denote the probabilities of fluctuations $u(d)
\rightarrow s + K^{-(0)}$, $u(d,s) \rightarrow u(d,s) + \eta$, and
$u(d,s) \rightarrow u(d,s) + \eta^{'}$. These fluctuation
parameters provide the basis to understand the extent to which the
quark sea contributes to the structure of the nucleon.

The GB field can be expressed in terms of the quark contents of
the GBs and their transition probabilities as \be {\Phi'} = \left(
\ba{ccc} \phi_{uu} u \bar u+ \phi_{ud} d \bar d +\phi_{us} s \bar
s& \varphi_{ud} u \bar d & \varphi_{us} u \bar s
\\ \varphi_{du} d \bar u & \phi_{du}u \bar u+ \phi_{dd} d \bar d
+\phi_{ds} s \bar s & \varphi_{ds} d \bar s
\\ \varphi_{su} s \bar u & \phi_{sd} s \bar d & \phi_{su} u \bar u
+ \phi_{sd} d \bar d +\phi_{ss} s \bar s \\ \ea \right),
\ee where \bea \phi_{uu} &=& \phi_{dd}= \frac{1}{2}
+\frac{\beta}{6} + \frac{\zeta}{3}\,,~~~~~ \phi_{ss}
=\frac{2\beta}{3} + \frac{\zeta}{3}\,,~~~~~ \phi_{us} = \phi_{ds}=
\phi_{su}=\phi_{sd}= -\frac{\beta}{3} + \frac{\zeta}{3}\,,
\nonumber \\ \phi_{du} &=& \phi_{ud}= -\frac{1}{2}
+\frac{\beta}{6} + \frac{\zeta}{3}\,,~~~~~ \varphi_{ud} =
\varphi_{du} = 1\,, ~~~~\varphi_{us} = \varphi_{ds}=\varphi_{su}=
\varphi_{sd}= \alpha \,.  \eea The contributions of the quark sea
coming from the fluctuation process in Eq. (\ref{basic}) can be
calculated  by substituting for every constituent quark $q \to
\sum P_q q + |\psi(q)|^2$, where $\sum P_q$ is the transition
probability of the emission of a GB from any of the $q$ quark and
$|\psi(q)|^2$ is the transition probability of the $q$ quark.

Before proceeding further, we briefly discuss the calculation of
the scalar matrix elements of the nucleon. The flavor structure of
the nucleon is defined as\cite{cheng1,cheng2,cheng3,cheng4} \be
\hat N \equiv \langle N |q \bar q| N \rangle, \label{bnb} \ee
where $|N\rangle$ is the nucleon wavefunction (detailed in
Ref.\cite{yaoubook}) and $q {\bar q}$ is the number operator for
the scalar quark content measuring the sum of the quark and
antiquark numbers \be q \bar q= \sum_{q=u,d,s} (n_q q + n_{\bar q
}{\bar q})= n_{u}u + n_{{\bar u}}{\bar u} + n_{d}d + n_{{\bar
d}}{\bar d} + n_{s}s + n_{{\bar s}}{\bar s}\,, \label{number} \ee
$n_{q({\bar q})}$ being the number of $q({\bar q})$ quarks. The
modified flavor structure of proton after the inclusion of the
effects of chiral fluctuations in the $\chi$CQM is expressed as
\be 2 P_u u + P_d d + 2 |\psi(u)|^2 + |\psi(d)|^2 \,, \ee where
the total probability of no emission of GB from a $q$ quark
($q=u,~d,~ s$) can be calculated from the Lagrangian and is given
by
\be
P_q=1-\sum P_q, \label{probability} \ee with \bea \sum P_u
&=&a(\phi^2_{uu}+\phi^2_{ud}+\phi^2_{us}+\varphi^2_{ud}+
\varphi^2_{us})\,,\label{probu} \\ \sum P_d
&=&a(\phi^2_{du}+\phi^2_{dd}+\phi^2_{ds}+\varphi^2_{du}+
\varphi^2_{ds})\,,\label{probd} \\ \sum P_s &=& a
\left(\phi^2_{su} + \phi^2_{sd} + \phi^2_{ss}+ \varphi^2_{su} +
\varphi^2_{sd} \right)\,, \label{probs} \\ |\psi(u)|^2 &=&
a\left[(2\phi^2_{uu}+\phi^2_{ud}+\phi^2_{us}+
\varphi^2_{ud}+\varphi^2_{us}){ u}+\phi^2_{uu}{ {\bar u}} \right.
\nonumber\\ &&\left.+(\phi^2_{ud}+\varphi^2_{ud})({d}+{ {\bar
d}})+ (\phi^2_{us}+\varphi^2_{us})({ s}+{ {\bar s}})\right ]\,,
\label{eqpsiu} \\ |\psi(d)|^2 &=& a\left
[(\phi^2_{du}+2\phi^2_{dd}+\phi^2_{ds}+
\varphi^2_{du}+\varphi^2_{ds}){ d}+\phi^2_{dd}{ {\bar d}}\right.
\nonumber \\ && \left. + (\phi^2_{du}+\varphi^2_{du})({ u}+{ {\bar
u}})+ (\phi^2_{ds}+\varphi^2_{ds})({ s}+{ {\bar s}})\right ]\,,
\label{eqpsid}\\ |\psi(s)|^2 &=& a\left[ \left(\phi^2_{su} +
\phi^2_{sd} + 2\phi^2_{ss}+ \varphi^2_{su} + \varphi^2_{sd}
\right)s + \phi^2_{ss}\bar s \right. \nonumber
\\ && \left.  +(\phi^2_{su} + \varphi^2_{su})(u + \bar u) + (\phi^2_{sd}
+ \varphi^2_{sd})(d + \bar d) \right]\,. \label{eqpsis} \eea

In terms of the transition probabilities, the `averaged' integrals
of the quark distribution functions are expressed as \be u-\bar u
= 2 \,,~~~~~ d-\bar d=1\,,~~~~~s-\bar s=0 \,, \ee where the
antiquark distribution functions are \bea \bar u &=&
a(2\phi_{uu}^2 + \phi_{du}^2 + \varphi_{du}^2)\,, \nonumber\\ \bar
d &=& a(2\phi_{ud}^2 +2\varphi_{ud}^2 + \phi_{dd}^2)\,,
\nonumber\\ \bar s &=& a(2\phi_{us}^2 + 2\varphi_{us}^2 +
\phi_{ds}^2 + \varphi_{ds}^2)\,. \label{barq} \eea


The pion-nucleon sigma term ($\sigma_{\pi N}$) affected by the
contributions of the quark sea is expressed as \be \sigma_{\pi N}=
\hat{m} \langle N| {\bar{u}u} + {\bar{d}d}|N \rangle= \hat{m}
\left( 3 + 6a \left(\phi^2_{uu} + \phi^2_{ud} + \varphi^2_{ud}
\right) \right) \,, \label{pion nucleon} \ee where ${\hat
m}=\frac{(m_u +m_d)}{2}$ is the average value of current $u$ and
$d$ quark masses evaluated at fixed gauge coupling and $q {\bar
q}$ is the scalar quark content\cite{sum}. Since $\sigma_{\pi N}$
provides restriction on the contribution of strange quarks in the
nucleon, it can be rewritten in terms of the strangeness content
in nucleon $y_N$ as \bea \sigma_{\pi N} = \hat m \frac{ \langle N|
{\bar{u}u} + {\bar{d}d} -2 {\bar{s}s} | N \rangle}{1- 2y_N}
=\frac{\hat {\sigma}}{1-2 y_N} \label{fs1} \,, \eea where we have
defined \be \hat {\sigma} = {\hat m}{\langle N| {\bar{u}u} +
{\bar{d}d} -2 {\bar{s}s} | N \rangle} = {\hat m} (3 +
6a(\phi^2_{uu} +\phi^2_{ud} + \varphi^2_{ud} -2 \phi^2_{us} -2
\varphi^2_{us} ) )\,,\ee  and \bea y_N &=& \frac{\langle
N|{\bar{s}} s |N \rangle} {\langle N| {\bar{u}u} + {\bar{d}d} | N
\rangle} = \frac{2a ( \phi_{us}^2 + \varphi_{us}^2)}{1 + 2 a
(\phi^2_{uu} +\phi^2_{ud} + \phi^2_{us} + \varphi^2_{ud}
+\varphi^2_{us}) }\,. \eea

Since $\sigma_{\pi N}$ is related to the hadron mass spectrum as
well as the quark mass ratio, therefore, following Ref.\cite{cheng
rev}, we can express $\hat {\sigma}$ as \be \hat {\sigma} =-
\frac{3(M_{\Xi} - M_{\Lambda})}{(1 - \frac{m_s}{\hat m})} \,, \ee
where $M_{\Xi}$ and $M_{\Lambda}$ are the baryon masses. The
latest accepted quark mass ratio $\frac{m_s}{\hat m}$ has the
value 22-30\cite{pdg}.

It is also important to define the strangeness fraction of the
nucleon which is related to the strangeness content in nucleon
$y_N$ as \be f_s = \frac{\langle N|{\bar{s}} s |N \rangle}
{\langle N| {\bar{u}u} + {\bar{d}d} + {\bar{s}s} | N \rangle}
=\frac{y_N}{1-y_N} \,. \label{fs} \ee This can further be
rewritten in terms of $\sigma_{\pi N}$  and $\hat {\sigma}$ and is
expressed as \bea f_s =
\frac{\sigma_{\pi N} - \hat{\sigma}}{3 \sigma_{\pi N} -
\hat{\sigma}} \,.\eea Another important parameter which is
completely determined from the strangeness content in nucleon
$y_N$ and the mass ratio is the strangeness sigma term \be
\sigma_s=m_s {\langle N| \bar{s}s | N \rangle} =\frac{1}{2}y_N
\frac{m_s}{\hat m} \sigma_{\pi N}\,. \ee According to NQM, the
valence quark structure of the nucleon does not involve strange
quarks. The validity of OZI rule\cite{ozi1,ozi2,ozi3,ozi4,ozi5} in
this case would imply $y_N=f_s=0$ or ${\hat \sigma}=\sigma_{\pi
N}$. For $\frac{m_s}{\hat m}=22$, the value of $\sigma_{\pi N}$
comes out to be close to 28 $MeV$. However, the most recent
analysis of experimental data gives higher values of $\sigma_{\pi
N}$ which points towards a significant strangeness content in the
nucleon.

Further, we can calculate the sigma terms corresponding to the
strange mesons. For example, the kaon-nucleon sigma term can be
expressed in terms of the scalar quark content and $\sigma_{\pi
N}$ as \be \sigma_{K N}= \frac{\sigma^u_{K N}+\sigma^d_{K N}}{2} =
\frac{\hat{m} + m_s}{2}\langle N| {\bar{u}u} + {\bar{d}d}+2
{\bar{s}s}|N\rangle
=\frac{\hat m +m_s}{4 \hat m}(2 \sigma_{\pi N} - \hat{\sigma}) \,,
\ee where $\sigma^u_{K N} = \frac{\hat{m} + m_s}{2}\langle N|
{\bar{u}u} + {\bar{s}s}|N\rangle $ and $\sigma^d_{K N} =
\frac{\hat{m} + m_s}{2}\langle N| {\bar{d}d} +
{\bar{s}s}|N\rangle$. Similarly,  the $\eta$-nucleon sigma term
can be expressed as  \bea \sigma_{\eta N} &=& \frac{1}{3} \langle
N|\hat{m}({\bar{u}u}+{\bar{d}d})+ 2 m_s
{\bar{s}s}|N\rangle  
= \frac{1}{3} \hat{\sigma}+ \frac{2 (m_s+ {\hat m})}{3 {\hat m
}}y_N \sigma_{\pi N} \,. \eea

In Table \ref{strange}, we have presented the results of our
calculations in $\chi$CQM pertaining to the scalar matrix elements
which are affected by the strangeness content of the nucleon as
well as the quantities which are affected by the quark mass ratio
as well as the strangeness content, for example, $\hat {\sigma}$,
$\sigma_s$, $\sigma_{\pi N}$, $\sigma_{K N}$, and $\sigma_{\eta
N}$. For the sake of comparison, we have also given the
corresponding quantities in NQM and the available phenomenological
values. To understand the implications of the strange quark mass
and SU(3) symmetry breaking, we have presented the results with
and without SU(3) symmetry breaking. A closer look at the
expressions  of these quantities reveals that the constant factors
represent the NQM results which do not include the effects of
chiral symmetry breaking. On the other hand, the factors with
transition probability $a$ represent the contribution from the
``quark sea'' in general (with or without SU(3) symmetry
breaking). As discussed earlier, the terms $\alpha$, $\beta$ and
$\zeta$ give the  SU(3) symmetry breaking effects. In the present
case, we have considered $a=0.12$, $\zeta=-0.15$,
$\alpha=\beta=0.45$ for the SU(3) symmetry breaking case whereas
under the SU(3) symmetric assumption we have taken
$\alpha=\beta=-\zeta=1$. Since the $\sigma$ terms are
characterized by the parameters of $\chi$CQM as well as the light
quark mass ratio, we have used the same set of parameters for
$\chi$CQM as discussed above and for $\frac{m_s}{\hat m}$, we have
used the most widely accepted value $\frac{m_s}{\hat m}=22$ from
the range $22-30$\cite{pdg}.

From the Table one finds that the present result for the
strangeness content in the nucleon $y_N$ and strangeness fraction
of the nucleon $f_s$ looks to be in agreement with the most recent
phenomenological results available which the NQM is unable to
explain. The non-zero values for   $y_N$ and $f_s$ in the present
case indicate that the chiral symmetry breaking is essential to
understand the significant role played by the quark sea. It is
also clear from the table that, in general, the quantities
involving the strange quark content are very sensitive to SU(3)
symmetry breaking. For example, the values of the strangeness
dependent quantities $y_N$ and  $f_s$ change to a large extent
when compared for the SU(3) symmetric and SU(3) symmetry breaking
case. The results for other quantities which do not have
strangeness contribution are not much different for both the
cases. The SU(3) symmetric results for $y_N$ and $f_s$ are $\sim
5-6$ times higher than the SU(3) symmetry breaking case. Such a
large value cannot be justified which is also in agreement with
the observations of other
authors\cite{gasser1,gasser2,cheng1,cheng2,cheng3,cheng4}.

A closer examination of the results reveals several interesting
points. We find that the $\sigma$ terms increase by taking lower
values of the quark mass ratio but it has been argued that the
possibility of readjusting the quark mass ratio to get higher
value of $\sigma$ term is ruled out\cite{domin}. For $\sigma_{\pi
N}$, the value of $\chi$CQM with SU(3) symmetry can give a value
in the higher range by adopting a larger value of $\hat {\sigma}$
however, as has been shown in our earlier work, SU(3) symmetry
does not give a satisfactory description of quark sea asymmetry
and spin related quantities. Also, the $\sigma_{K N}$ and
$\sigma_{\eta N}$ become strangely large for the SU(3) symmetric
case which confirms that SU(3) symmetry breaking effects should be
taken into account. A refinement in the analysis of $\pi-N$
scattering giving higher values of $\sigma_{\pi N}$ would not only
strengthen the mechanism of chiral symmetry breaking generating
the appropriate amount of strangeness in the nucleon but would
also justify the consequences of SU(3) symmetry breaking
mechanism. The $\sigma_{K N}$ and $\sigma_{\eta N}$ terms are
found to be quite sensitive to $y_N$.

The effects of external effects like external magnetic and
external gravitational field can easily be incorporated into the
calculations of $\chi$CQM. Without getting into the details, our
results show that the effects of magnetic field and gravitational
field contribute towards $\chi$SB in the opposite directions. This
is in agreement with the results in Ref.\cite{muta}. The chiral
symmetry is broken by the presence of magnetic field whereas the
presence of gravitational field tends to restore chiral symmetry.
This leads to a very small overall contribution and the exact
order of magnitude can be estimated in a more detailed
calculation.

Future DA$\Phi$NE experiments\cite{dafne} will allow a
determination of KN sigma terms and hence could restrict the model
parameters and provide better knowledge of strangeness content of
the nucleon. Further, it would be interesting to find out the role
of strangeness content in the nucleon in the hyperon-antihyperon
production in heavy ion collisions.


To summarize, the  $\chi$CQM is able to phenomenologically
estimate the quantities having implications for chiral symmetry
breaking. In particular, it provides a fairly good description of
the scalar matrix elements having implications for hidden
strangeness component in the nucleon, for example, the strangeness
content in the nucleon $y_N$ and strangeness fraction of the
nucleon $f_s$. The non-zero values for $y_N$ and $f_s$ indicate
that the chiral symmetry breaking is essential to understand the
significant role of non-valence quarks in the nucleon structure.
The significant contribution of the strangeness is consistent with
the recent available experimental results which justify that
chiral symmetry breaking and SU(3) symmetry breaking play an
important role in understanding the flavor structure of the
nucleon.

The calculations have also been extended to predict the
meson-nucleon sigma terms ($\sigma_{KN}$ and $\sigma_{\eta N}$).
The future DA$\Phi$NE experiments to determine KN sigma terms
could restrict the model parameters and provide better knowledge
of strangeness content of the nucleon. The role of strangeness
content in the nucleon would also have important implications for
the  hyperon-antihyperon production in the heavy ion collisions.

In conclusion, we would like to state that chiral symmetry
breaking is the key to understand the hidden strangeness content
of the nucleon. In the nonperturbative regime of QCD, constituent
quarks and the weakly interacting Goldstone bosons constitute the
appropriate degrees of freedom at the leading order.


\vskip .2cm {\bf ACKNOWLEDGMENTS}\\ The work of H.D. was partly
supported by Department of Science and Technology, Government of
India through grant no. SR/S2/HEP-0028/2008.

\begin{table}[h]
\tbl{The $\chi$CQM results for the scalar matrix elements of the
nucleon and the meson-nucleon sigma terms.}
{\begin{tabular}{@{}ccccc@{}} \toprule
 &  & NQM\cite{nqm1,nqm2,nqm3,nqm4,nqm5} & $\chi$CQM & $\chi$CQM
\\ Quantity& Phenomenology &  & with SU(3)& with SU(3)
\\& & & symmetry &symmetry breaking \\ \colrule $\langle N |{\bar
u} u| N \rangle$ &...&$ \leq $ 2&2.41&2.44 \\ $\langle N |{\bar d}
d| N \rangle$ &...&$ \leq $ 1&1.75&1.68\\ $\langle N |{\bar s} s|
N \rangle$ &...& 0.0&1.08&0.18\\ $y_N$ & $0.11 \pm
0.07$\cite{gasser1,gasser2}& 0.0 & 0.26 &0.044 \\ $f_s$ &$0.10 \pm
0.06$\cite{nutev}& 0.0 & 0.21 & 0.042\\ ${\hat \sigma}$ & ...&
28.57& 28.57&28.57 \\ $\sigma_{s}$ &...&0 &168.71&15.12 \\
$\sigma_{\pi N}$  &...&28.57&59.25&31.32 \\ $\sigma_{K N}$
&...&164.29&517.04&195.90\\ $\sigma_{\eta N}$  &...&9.52&
244.70&30.60 \\ \botrule
\end{tabular}
\label{strange}}
\end{table}

\end{document}